\documentclass[aps,prd,preprint,showkeys,showpacs,unsortedaddress,superscriptaddress]{revtex4-1}
\usepackage{amsmath, amsfonts, amssymb, mathrsfs,float}
\usepackage{slashbox}
\usepackage{graphicx}
\usepackage{bm,color}
\usepackage{multirow}
\usepackage{caption,subcaption}
\usepackage{epsfig}
\usepackage{epstopdf}
\usepackage{epsf}
\newcommand{\nn}{\nonumber}

\allowdisplaybreaks

\begin{document}

\title{The quasi-Keplerian motion in the regular Bardeen spacetime}

\author{Jie Li}%
\affiliation{School of Nuclear Science and Technology, University of South China, Hengyang, 421001, China}

\author{Bo Yang}
\email{yb@usc.edu.cn.}
\affiliation{School of Mathematics and Physics, University of South China, Hengyang, 421001, China}

\author{Wenbin Lin}%
\email{lwb@usc.edu.cn.}
\affiliation{School of Mathematics and Physics, University of South China, Hengyang, 421001, China}
\affiliation{School of Physical Science and Technology, Southwest Jiaotong University, Chengdu, 610031, China}

\date{\today}

\begin{abstract}
The second post-Newtonian solution for the quasi-Keplerian motion of a test particle in the gravitational field of regular Bardeen black hole is derived. The solution is formulated in terms of the test particle's orbital energy and angular momentum, as well as the mass and magnetic charge of the Bardeen black hole. The leading effects of the magnetic charge on the test particle's orbit and motion including perihelion precession are displayed explicitly. In particular, it is shown that to the second post-Newtonian order the magnetic charge does not affect the test particle's orbital period. 
\end{abstract}

\maketitle

%
%
%
%
%
%
%
%
%
%
%
\section{introduction}
Black hole is one of the most prominent predictions of general relativity. With the development of science and technology, the gravitational wave generated by the collision of two black holes was detected~\cite{Abbott2016,Cervantes2016,Krolak2021} and the shadow of black holes was revealed by Event Horizon Telescope collaboration~\cite{Akiyama2019}, which directly confirmed the existence of black holes, greatly stimulated people's interest in the physics of black holes. However, spacetime singularity is one of the most intriguing problems. According to Penrose and Hawking's singularity theorems, every black hole inevitably contains a singularity within the framework of Einstein's general relativity as long as some certain conditions are satisfied~\cite{Hawking1970}. At the singularity of spacetime, physical quantities and geometry diverge, and the laws of physics lose their power completely.
To avoid the singularity problem, Bardeen~\cite{Bardeen1968} first proposed a regular black hole model named the Badeen black hole, which was later proved to be the exact solution of the Einstein's field equations coupling with nonlinear electrodynamics~\cite{AyonBeato2000}. Subsequently, a number of other solutions to regular black holes such as Hayward black hole~\cite{Hayward2006}, Ay\'{o}n-Beato and Garc\'{\i}a black hole~\cite{AyonBeato1998} and  Simpson-Visser black holes~\cite{Simpson2019} have been proposed, and the rotating solutions of these black holes has been obtained in~\cite{Toshmatov2014,Abdujabbarov2016,Mazza2021}.

Up to now, Bardeen black hole has attracted extensive attention. Bret\'{o}n et al studied the stability of black holes with nonlinear electromagnetic fields~\cite{Breton2015} and Quasinormal modes was investigated by Fernando and Correa~\cite{Fernando2012}. Gravitational lensing by black holes with nonlinear electromagnetic fields was investigated in~\cite{Eiroa2011,Ghaffarnejad2016}. By analyzing the behavior of the effective potential, Zhou et al investigated the time-like and null geodesic structures~\cite{Zhou2012}. Abdujabbarov et al studied the Shadow of rotating regular black holes, and found that magnetic charge decreases the size of the shadow~\cite{Abdujabbarov2016}. Gao et al studied the bound orbits around the Badeen black hole~\cite{Gao2020} and greybody factor are studied in~\cite{Sharif2019,AmaTulMughani2021,Sharif2022}. The thermodynamic properties of the Bardeen black holes are discussed in~\cite{Man2014}. Some properties of Bardeen black holes in other modified theories have also been investigated~\cite{Mehdipour2016,Ghaderi2018,Saleh2018,Saleh20182,Pourhassan2019,SinghDV2020,SinghBK2020,Kumar2022}.

The motion of bodies in the vicinity of black holes is an important problem of relativistic astrophysics. 
To solve the motion of the bodies in the strong gravitational fields, the post-Newtonian (PN) approximations have been extensively applied, and various analytical solutions have been obtained, including the first and higher PN effects of the mass~\cite{Brumberg1972,Soffel1987,Soffel1989,KlionerKopeikin1994,KopeikinEfroimskyKaplan2012,DamourSchafer1988,SchaferWex1993,MGS2004,Boetzel2017,Cho2018,SoffelHan2019,YangLin2020a,YangLin2020b}, the leading order~(1.5PN) spin effects~\cite{Wex1995,GPV1998a,GPV1998b,GPV1998c,KonigsdorfferGopakumar2005,KMG2005,KonigsdorfferGopakumar2006b,GopakumarSchafer2011,BMFB2013,GergelyKeresztes2015,Mikoczi2017} on the general motion, as well as the leading order~(2PN) mass quadrupole effects on the circular motion~\cite{Poisson1998}.

In our previous work~\cite{YangLin2019,YangJiangLin2022}, we have derived the quasi-Keplerian motion for the neutral test particle in the Reissner-Nordstr\"{o}m spacetime, and that for the charged test particle. We obtained the effects of the spin-induced quadrupole on the equatorial motion of a test particle in the Kerr spacetime~\cite{YangLin2020c} and the next-to-leading order spin-orbit effects on the test particle's generally inclined motion~\cite{YangLin2021}. In this work, we will derive the quasi-Keplerian motion of a test particle in the gravitational field of regular Bardeen black hole under the standard spherical coordinates. The magnetic charge effects of the Bardeen black hole on the quasi-Keplerian motion including the orbital perihelion precession are displayed explicitly.

The rest of this paper is organized as follows. Section \ref{sec:2nd} introduces the standard metric of Bardeen black hole in the 2PN approximation, the corresponding Lagrangian, as well as the orbital energy and angular momentum. In Section \ref{sec:3rd} we present the detail derivation on the 2PN solution for the quasi-Keplerian motion. Summary is given in section \ref{sec:4th}.

\section{The 2PN Lagrangian, energy, angular momentum in Bardeen spacetime}\label{sec:2nd}

\begin{figure}[h!]
\centering
\begin{subfigure}[t]{0.575\textwidth}
\centering
\includegraphics[width=9cm,height = 5.9cm]{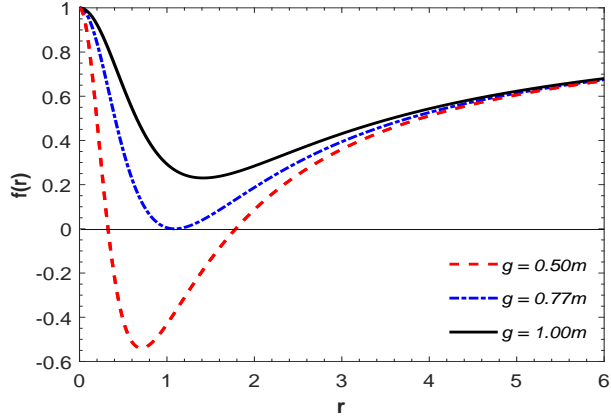}
\end{subfigure}
\caption{ Horizons of the Bardeen black hole with different values of $g$.}~\label{Fig1}
\end{figure}

Assume the regular Bardeen black hole is located at the coordinate origin. In the standard spherical coordinates, the exact metric of Bardeen black hole can be 
expressed as
\begin{eqnarray}
&& ds^{2}=-f(r)dt^{2}+f(r)^{-1}dr^2+r^2d\theta^2+r^2\sin \theta^2 d\phi^2,  \label{HarmonicKerr}
\end{eqnarray}
where the lapse function $f(r)$ is given by
\begin{eqnarray}
&& f(r)=1-\frac{2mr^2}{(r^2+g^2)^{3/2}}~,
\end{eqnarray}
here $m$ denotes the standard gravitational mass parameter, while $g$ corresponds to the magnetic charge parameter measured in units of $m$. When $g=0$, this metric behaves as a Schwarzschild metric. When $g^2\leq\frac{16}{27}m^2$, a no-singularity black hole is described. In particular, when $g^2=\frac{16}{27}m^2$, single horizon solution is allowed, i.e., regular extremal black hole as shown in Fig~\ref{Fig1}. The gravitational constant and the speed of light in vacuum are set as 1.

According to the theory of the PN approximation, this metric can be expanded into the power of $\frac{m}{r}$ with $r\!=\!|\bm{x}|$. Keeping all terms to the 2PN order for the dynamics of the massive particle, we have
\begin{eqnarray}
&& g_{00}= -1 +\frac{2m}{r}-\frac{3 g^2m}{r^3}~,\label{eq:metric-1nd}\\
&& g_{0i}= 0~,\label{eq:metric-2nd}\\ 
&& g_{ij}=\delta_{ij}+\Big(\frac{2m}{r}+\frac{4m^2}{r^2}\Big)\frac{x^ix^j}{r^2} ,\label{eq:metric-3nd}
\end{eqnarray}
where $i$ and $j$ running from 1 to 3, $\epsilon_{ijk}$ is Levi-Civita symbol~\cite{Weinberg1972}.

With the 2PN metric Eqs.\,(\ref{eq:metric-1nd})-(\ref{eq:metric-3nd}), the Lagrangian of the test particle can be written as
\begin{eqnarray}
&& {\rm L} = \frac{1}{2}\bm{v}^{2}+\frac{m}{r}+\frac{1}{8}\bm{v}^{4}
+\frac{1}{2}\frac{m}{r}\bm{v}^{2}+\frac{1}{2}\frac{m^2}{r^2}+\frac{m(\bm{v}\!\cdot\!\bm{x})^2}{r^3}+\frac{1}{2}\frac{m^3}{r^3}\Big(1-\frac{3g^2}{m^2}\Big)\nn\\
&&~~~~~~+\frac{1}{16}\bm{v}^{6}+\frac{3}{4}\frac{m^2}{r^2}\bm{v}^{2}+\frac{3}{8}\frac{m}{r}\bm{v}^{4}
+\frac{3m^2(\bm{v}\!\cdot\!\bm{x})^2}{r^4}+\frac{m(\bm{v}\!\cdot\!\bm{x})^2\bm{v}^{2}}{2r^3}~,\label{eq:Lagrangian}
\end{eqnarray}
where $\bm{v}$ denotes the the test particle's velocity.
According to this Lagrangian, the energy $\mathcal{E}$ and the angular momentum $\mathcal{J}$ of the equatorial motion can be obtained as
\begin{eqnarray}
&& \mathcal {E}
=\frac{1}{2}\bm{v}^{2}-\frac{m}{r}+\frac{3}{8}\bm{v}^{4}+\frac{1}{2}\frac{m}{r}\bm{v}^{2}-\frac{1}{2}\frac{m^2}{r^2}+\frac{m(\bm{v}\!\cdot\!\bm{x})^2}{r^3}
-\frac{1}{2}\frac{m^3}{r^3}\Big(1-\frac{3g^2}{m^2}\Big)\!\nn\\
&& ~~~~~~+\frac{5}{16}\bm{v}^{6}+
\frac{3}{4}\frac{m^2}{r^2}\bm{v}^{2}+\frac{9}{8}\frac{m}{r}\bm{v}^{4}
+\frac{3m^2(\bm{v}\!\cdot\!\bm{x})^2}{r^4}+\frac{3}{2}\frac{m(\bm{v}\!\cdot\!\bm{x})^2\bm{v}^{2}}{r^3}~,\label{eq:total-energy}\\
&& \mathcal{J}
= |\bm{x}\!\times\!\bm{v}|\Big[1+\frac{1}{2}\bm{v}^{2}
+\frac{m}{r}+\frac{3}{8}\bm{v}^{4}+\frac{3}{2}\frac{m^2}{r^2}+\frac{3}{2}\frac{m}{r}\bm{v}^{2}+\frac{m(\bm{v}\!\cdot\!\bm{x})^2}{r^3}\Big]~.
\label{eq:total-angular momentum}
\end{eqnarray}

\section{The quasi-Keplerian motion in the 2PN approximation}\label{sec:3rd}

We deduce the analytical solution for the quasi-Keplerian motion according to the same procedure given by Brumberg~\cite{Brumberg1972}.

The trajectory of the test particle in the equatorial plane of the Bardeen black hole can be expressed as
\begin{equation}
\bm{x} = r(\cos\phi\,\bm{e}_{x}+\sin\phi\,\bm{e}_{y})~,\label{eq:x1}
\end{equation}
where $\phi$ is the azimuthal angle. $\bm{e}_{x}$ and $\bm{e}_{y}$ are the unit vectors of the $x$-axis and $y$-axis.

We can re-write the orbital energy and angular momentum in Eqs.~(\ref{eq:total-energy})-(\ref{eq:total-angular momentum}) as
\begin{eqnarray}
&& \hskip -0cm \mathcal {E}=
\frac{1}{2}(\dot{r}^2\!+\!r^2\dot{\phi}^2)\!-\!\frac{m}{r}\!+\!\frac{3}{8}(\dot{r}^2\!+\!r^2\dot{\phi}^2)^2\!+\!\frac{1}{2}\frac{m}{r}(\dot{r}^2\!+\!r^2\dot{\phi}^2)
\!-\!\frac{m^2}{2r^2}\!+\!\frac{m}{r}\dot{r}^2\!-\!\frac{m^3}{2r^3}\Big(1\!-\!\frac{3g^2}{m^2}\Big)\nn\\
&& \hskip 1cm +\frac{5}{16}(\dot{r}^2\!+\!r^2\dot{\phi}^2)^3\!+\!\frac{3}{4}\frac{m^2}{r^2}(\dot{r}^2\!+\!r^2\dot{\phi}^2)\!
+\!\frac{9}{8}\frac{m}{r}(\dot{r}^2\!+\!r^2\dot{\phi}^2)^2
\!+\!\frac{3m^2}{r^2}\dot{r}^2\!+\!\frac{3}{2}\frac{m}{r}\dot{r}^2(\dot{r}^2\!+\!r^2\dot{\phi}^2) ,\label{eq:Brumberg-total-energy}\\
&& \hskip -0cm\mathcal{J}=r^2\dot{\phi}
\Big[1\!+\!\frac{1}{2}(\dot{r}^2\!+\!r^2\dot{\phi}^2)\!+\!\frac{m}{r}\!+\!\frac{3}{8}(\dot{r}^2\!+\!r^2\dot{\phi}^2)^2\!+\!
\frac{3}{2}\frac{m}{r}(\dot{r}^2\!+\!r^2\dot{\phi}^2)\!+\!\frac{3}{2}\frac{m^2}{r^2}\!+\!\frac{m}{r}\dot{r}^2\Big]~,\label{eq:Brumberg-total-angular momentum}
\end{eqnarray}
where the dot means the derivative with respect to the time. From these two expressions we can obtain
\begin{equation}
r^4\dot{\phi}^2=\mathcal{J}^2\Big(1\!-\!2\,\mathcal {E}\!-\!\frac{4m}{r}\!+\!3\,\mathcal {E}^2\!+\!\frac{4m^2}{r^2}\!+\!8\,\mathcal {E}\frac{m}{r}\Big)~,\label{eq:r4dottheta}
\end{equation}
and
\begin{equation}
 \dot{r}^2=A+\frac{B}{r}+\frac{C}{r^2}+\frac{D}{r^3}+\frac{E}{r^4}~,\label{eq:rdot20}
\end{equation}
with
\begin{eqnarray}
&& A=2\mathcal {E}\Big(1\!-\!\frac{3}{2}\,\mathcal {E}\!+\!2\,\mathcal {E}^2\Big)~,\\
&& B=2m\Big(1\!-\!6\,\mathcal {E}\!+\!9\,\mathcal {E}^2\Big)~,\\
&& C=-\mathcal{J}^2\Big(1\!-\!2\,\mathcal{E}\!+\!\frac{8m^2}{\mathcal{J}^2}\!-\!24\frac{m^2\mathcal {E}}{\mathcal{J}^2}\!+\!3\,\mathcal {E}^2\Big)~,\\
&& D=6m \mathcal{J}^2\Big[1\!-\!2\,\mathcal{E}\!+\!\frac{4}{3}\frac{m^2}{\mathcal{J}^2}\Big(1\!-\!\frac{3g^2}{8m^2}\Big)\Big]~,\\
&& E=-12m^2 \mathcal{J}^2~.
\end{eqnarray}
Throughout this paper, the signs ``$\pm$" denote the anti-clockwise motion and the clockwise motion, respectively. 

utilizing the relation
\begin{equation}
 \dot{r}^2=\Big[\frac{d(1/r)}{d\phi}\Big]^2(r^4\dot{\phi}^2)~,\label{eq:rdot2-relation}
\end{equation}
and plugging Eqs.~(\ref{eq:r4dottheta})-(\ref{eq:rdot20})  into (\ref{eq:rdot2-relation}), we can
express the radial equation in the form
\begin{eqnarray}
&& \Big[\frac{d(1/r)}{d\phi}\Big]^2=A'+\frac{B'}{r}+\frac{C'}{r^2}+\frac{D'}{r^3}+\frac{E'}{r^4}~,\label{eq:the-radial-equation1}
\end{eqnarray}
with
\begin{eqnarray}
&& A'=\frac{2\mathcal {E}}{\mathcal{J}^2}\Big(1\!+\!\frac{1}{2}\,\mathcal {E}\Big)~,\\
&& B'=\frac{2m}{\mathcal{J}^2}~,\\
&& C'=-1~,\\
&& D'=2m\Big(1\!-\!\frac{3g^2}{2\mathcal{J}^2}\Big)~,\\
&& E'=0~.
\end{eqnarray}
Since the right hand side of Eq.~(\ref{eq:the-radial-equation1}) is a fourth-order polynomial in $r^{-1}$, we can further re-write it as
\begin{eqnarray}
&& \Big[\frac{d(1/r)}{d\phi}\Big]^2=\Big[\frac{1}{r}\!-\!\frac{1}{a_{r}
(1+e_{r})}\Big]\Big[\frac{1}{a_{r}(1-e_{r})}\!-\!\frac{1}{r}\Big]\Big(C_{1}+\frac{C_{2}}{r}+\frac{C_{3}}{r^2}\Big)~.\label{eq:the-radial-equation2}
\end{eqnarray}
Comparing the coefficients between Eq.~(\ref{eq:the-radial-equation1}) and Eq.~(\ref{eq:the-radial-equation2}), we have 
\begin{eqnarray}
&& a_{r}=\frac{m}{-2\mathcal{E}}\Big[1\!+\!\frac{3}{2}\mathcal{E}\!+\!\frac{1}{4}\mathcal{E}^2\!+\!8\frac{m^2\mathcal{E}}{\mathcal{J}^2}\Big(1\!-\!\frac{3g^2}{8m^2}\Big)\Big]~,\label{eq:aPN_EJ}\\
&& e_r^2=1\!+\!\frac{2\mathcal{E}\mathcal{J}^2}{m^2}\!-\!\mathcal{E}\Big(8\!+\!7\frac{\mathcal{E}\mathcal{J}^2}{m^2}\Big)
\!+\!\,\mathcal{E}^2\Big[16\frac{\mathcal{E}\mathcal{J}^2}{m^2}\!-\!20\Big(1\!-\!\frac{3g^2}{5m^2}\Big)\!-\!32\frac{m^2}{\mathcal{E}\mathcal{J}^2}
\Big(1\!-\!\frac{3g^2}{8m^2}\Big)\Big]~,\label{eq:ePN_EJ}\\
&& C_{1}=1\!-\!\frac{4m^2}{\mathcal{J}^2}\!-\!8\frac{m^2\mathcal{E}}{\mathcal{J}^2}\!-\!\frac{16m^4}{\mathcal{J}^4}\Big(1\!-\!\frac{3g^2}{8m^2}\Big)~,\label{eq:C1}\\
&& C_{2}=-2m\Big(1\!-\!\frac{3g^2}{2\mathcal{J}^2}\Big)~,\label{eq:C2}\\
&& C_{3}=0~.\label{eq:C3}
\end{eqnarray}
It can be seen from Eq.~(\ref{eq:the-radial-equation2}) that $r_{\pm}=a_{r}(1\pm e_{r})$ represent the maximal and minimal values for $r$. Hence, $a_{r}$ and $e_{r}$ can be regarded as the semi-major axis and the eccentricity of the quasi-Keplerian orbit.

The solution of Eq.~(\ref{eq:the-radial-equation2}) can be written as:
\begin{eqnarray}
&& r=\frac{a_{r}(1-e_{r}^2)}{1+e_{r}\cos{f}}~,\label{eq:rBrumberg}
\end{eqnarray}
with $f$ being the true anomaly for the quasi-Keplerian orbit and satisfying
\begin{eqnarray}
&& \Big(\frac{df}{d\phi}\Big)^2=C_{1}+\frac{C_{2}}{r}+\frac{C_{3}}{r^2}~.\label{eq:f_phi2}
\end{eqnarray}

Substituting Eqs.~(\ref{eq:C1})-(\ref{eq:rBrumberg}) into Eq.~(\ref{eq:f_phi2}), we have
\begin{eqnarray}
&& \frac{df}{d\phi}=F\Big\{1\!-\! \frac{m^2}{\mathcal{J}^2}\Big[1\!+\!2\mathcal{E}\!+\!10\frac{m^2}{\mathcal{J}^2}\Big(1\!-\!\frac{3g^2}{20m^2}\Big)\Big]e_{r} \cos f
\!-\!\frac{m^4}{4\mathcal{J}^4}e_{r}^2 \cos 2f\Big\}~,\label{eq:dfdphi}
\end{eqnarray}
with
\begin{eqnarray}
F=\Big[1\!-\!\frac{3m^2}{\mathcal{J}^2}\!-\!\frac{13}{2}\frac{m^2\mathcal{E}}{\mathcal{J}^2}\!-\!\frac{67}{4}\frac{m^4}{\mathcal{J}^4}\Big(1\!-\!\frac{18g^2}{67m^2}\Big)
\Big]~.\label{eq:F}
\end{eqnarray}

Making integration of Eq.~(\ref{eq:dfdphi}), we can obtain 
\begin{equation}
\phi \Big(\frac{2\pi}{\Phi}\Big)=f+\frac{m^2}{\mathcal{J}^2}\Big[1\!+\!2\mathcal{E}\!+\!10\frac{m^2}{\mathcal{J}^2}\Big(1\!-\!\frac{3g^2}{20m^2}\Big)\Big]e_{r} \sin f\!+\!\frac{3}{8}\frac{m^4}{\mathcal{J}^4}e_r^2\sin 2f~,\label{eq:phi_f}
\end{equation}
with
\begin{equation}
 \Phi=2\pi\Big[1\!+\!3\frac{m^2}{\mathcal{J}^2}\!+\!\frac{15}{2}\frac{m^2\mathcal{E}}{\mathcal{J}^2}\!+\!\frac{105}{4}\frac{m^4}{\mathcal{J}^4}
\Big(1\!-\!\frac{6g^2}{35m^2}\Big)\Big]~.\label{eq:Phi}
\end{equation}

Finally, we derive the time dependence of the quasi-Keplerian motion.
Combining Eqs.~(\ref{eq:r4dottheta}) and (\ref{eq:dfdphi})-(\ref{eq:F}), we have
\begin{eqnarray}
&& \hskip -0cm r^2\dot{f}=\mathcal{J}\Big\{1\!-\!\mathcal {E}\!-\!2\frac{m}{r}\!-\!3\frac{m^2}{\mathcal{J}^2}
\!+\!\mathcal {E}^2\!+\!2\mathcal {E}\frac{m}{r}\!-\!\frac{7}{2}\frac{m^2\mathcal{E}}{\mathcal{J}^2}\!
+\!6\frac{m}{r}\frac{m^2}{\mathcal{J}^2}\!-\!\frac{m^4}{\mathcal{J}^4}
\Big(\frac{67}{4}\!-\!\frac{9g^2}{2m^2}\Big)\nn\\
&& \hskip 1.5cm -\frac{m^2}{\mathcal{J}^2}\Big[1\!+\!\mathcal {E}\!-\!2\frac{m}{r}\!+\!\frac{m^2}{\mathcal{J}^2}\Big(7\!-\!\frac{3g^2}{2m^2}\Big)\Big]e_{r} \cos f
\!-\!\frac{1}{4}\frac{M^4}{\mathcal{J}^4} e_{r}^2\cos 2f\Big\}.\label{eq:dfdt}
\end{eqnarray}

Introducing the post-Newtonian eccentric anomaly $u$ by the relations
\begin{equation}
\hskip -0cm \sin f\!=\!\frac{(1\!-\!e_{r}^2)^{\frac{1}{2}}\sin u}{1\!-\!e_{r}\cos u};\,\,\cos f\!=\!\frac{\cos u\!-\!e_{r}}{1\!-\!e_{r}\cos u};\,\,
 f\!=\!2 \!\arctan\!\Big(\sqrt{\frac{1\!+\!e_{r}}{1\!-\!e_{r}}}\tan \frac{u}{2}\Big),
\label{eq:u-relation}
\end{equation}
we have
\begin{eqnarray}
\frac{df}{dt}=\frac{(1-e_{r}^2)^{1/2}}{1-e_{r}\cos u}\frac{du}{dt}~,\label{eq:dfdu}
\end{eqnarray}
and we can formulate the orbit given in Eq.~(\ref{eq:rBrumberg}) in terms of $u$ as
\begin{equation}
 r=a_{r}(1-e_{r} \cos u)~.\label{eq:r}
\end{equation}

Integrating Eq.~(\ref{eq:dfdt}) and making use of Eqs.~(\ref{eq:u-relation})-(\ref{eq:r}), we can achieve the final piece of the 2PN closed-form solution for the equatorial motion in Bardeen spacetime.
\begin{equation}
 t\Big(\frac{2\pi}{{\rm T}_u}\Big)  = u-e_t \sin{u}+ \frac{30m \,\mathcal{E}^2 }{\sqrt{-2\,\mathcal{E}\mathcal{J}^2}} (f-u)~,\label{eq:nt}
\end{equation}
with ${\rm T}_{\!u}$ being the period for the eccentric anomaly $u$ of the quasi-Keplerian motion
\begin{equation}
 {\rm T}_u=\frac{2\pi m}{(-2\mathcal{E})^{\frac{3}{2}}}\Big[1\!-\!\frac{15}{4}\mathcal{E}\!-\!\frac{105}{32}\mathcal{E}^2\!+\!\frac{30m\mathcal{E}^2}{\sqrt{-2\mathcal{E}\mathcal{J}^2}}\Big]~,
\end{equation}
and $e_t$ being the time eccentricity
\begin{equation}
 e_t=e_r\Big[1\!+\!6\,\mathcal{E}\!+\!27\,\mathcal{E}^2\!-\!\frac{30m\mathcal{E}^2}{\sqrt{-2\mathcal{E}\mathcal{J}^2}}\!+\!8\frac{m^2\mathcal{E}}{\mathcal{J}^2}
\Big(1\!-\!\frac{3g^2}{8m^2}\Big)\Big]~.
\end{equation}

In the literature, one usually replace the true anomaly $f$ in the formula of quasi-Keplerian equation with another true anomaly $\upsilon$, which requires that $\sin \upsilon$ contribution vanishes at each PN order in $\phi (\frac{2\pi}{\Phi})$~\cite{MGS2004,KonigsdorfferGopakumar2005,THS2010}.

Following the same method given in Ref~\cite{THS2010}, we set
\begin{equation}
 \upsilon=2 \arctan \Big(\sqrt{\frac{1+e_{\phi}}{1-e_{\phi}}}\tan \frac{u}{2}\Big)~,\label{eq:upsilon_u}
\end{equation}
with
\begin{equation}
 e_\phi=e_r(1+\epsilon\, c_1+\epsilon^2\, c_2)~,
\end{equation}
differing from the radial eccentricity $e_r$ by some 1PN and 2PN level corrections $c_1$ and $c_2$. Here $\epsilon$ only denotes the PN order and does not have any value. Eliminating $u$ in Eq.~(\ref{eq:u-relation}) with the help of Eq.~(\ref{eq:upsilon_u}), we have~\cite{THS2010}
\begin{equation}
\hskip -0.05cm f\!=\!\upsilon+\epsilon\, c_1\frac{e_r}{e_r^2\!-\!1}\sin\upsilon
+\epsilon^2\,\Big[\Big(c_2\!-\!c_1^2\frac{e_r^2}{e_r^2\!-\!1}\Big)\frac{e_r}{e_r^2
\!-\!1}\sin \upsilon+\frac{c_1^2}{4}\frac{e_r^2}{(e_r^2\!-\!1)^2}\sin 2\upsilon\Big]. 
\end{equation}
Substituting this result into Eq.~(\ref{eq:phi_f}) and requiring the $\sin \upsilon$ term to vanish in $\phi (\frac{2\pi}{\Phi})$, we can obtain
\begin{eqnarray}
&& c_1=-2\mathcal{E}~,\\
&& c_2=-\!18\mathcal{E}\frac{m^2}{\mathcal{J}^2}\Big(1\!-\!\frac{g^2}{6m^2}\Big)~,
\end{eqnarray}
which lead to
\begin{equation}
 e_{\phi} = e_r \Big[1\!-\!2\mathcal{E}\!-\!18\frac{m^2\mathcal{E}}{\mathcal{J}^2}\Big(1\!-\!\frac{g^2}{6m^2}\Big)\Big]~,
\end{equation}
\begin{equation}
 \phi \Big(\frac{2\pi}{\Phi}\Big)= \upsilon \!+\!\frac{1}{8}\frac{m^4}{\mathcal{J}^4}e_r^2\sin 2\upsilon~.
\end{equation}

With the true anomaly $\upsilon$, we can re-express the time dependance of the quasi-Keplerian motion Eq.~(\ref{eq:nt}) in the form of
\begin{equation}
 t\Big(\frac{2\pi}{{\rm T}_u}\Big)  = u-e_t \sin{u}+\frac{30m \,\mathcal{E}^2 }{\sqrt{-2\,\mathcal{E}\mathcal{J}^2}} (\upsilon-u)~.
\end{equation}

Notice that all the formulas in this work are valid up to the 2PN accuracy.

\section{Summary}\label{sec:4th}
We start with the 2PN metric of the Bardeen black hole in the standard coordinates, calculate the corresponding orbital energy and angular momentum of the test particle in the equatorial plane, and then through an iterative method and function fitting method, derive the 2PN solution for the quasi-Keplerian motion in the Bardeen spacetime. We obtain two slightly different but equivalent formulations in the 2PN approximation.

The results are summarized as follows.

The first formulation can be expressed as
\begin{eqnarray}
&& \bm{x} = r(\cos\phi\,\bm{e}_{x}+\sin\phi\,\bm{e}_{y})~,\nn\\
&& r=a_{r}(1-e_{r} \cos u)~,\nn\\
&& \phi \Big(\frac{2\pi}{\Phi}\Big)=f+ N_0\sin f+N_1\sin 2f~,\nn\\
&& f=2 \arctan \Big(\sqrt{\frac{1+e_{r}}{1-e_{r}}}\tan \frac{u}{2}\Big)~,\nn\\
&&  t\Big(\frac{2\pi}{{\rm T}_u}\Big)  = u-e_t \sin{u}+ N_2 (f-u)~,\nn
\end{eqnarray}
and the second formulation can be expressed as
\begin{eqnarray}
&& \bm{x} = r(\cos\phi\,\bm{e}_{x}+\sin\phi\,\bm{e}_{y})~,\nn\\
&& r=a_{r}(1-e_{r} \cos u)~,\nn\\
&& \phi \Big(\frac{2\pi}{\Phi}\Big)= \upsilon + N_3 \sin 2\upsilon~,\nn\\
&& \upsilon=2 \arctan \Big(\sqrt{\frac{1+e_{\phi}}{1-e_{\phi}}}\tan \frac{u}{2}\Big)~,\nn\\
&&  t\Big(\frac{2\pi}{{\rm T}_u}\Big)  = u-e_t \sin{u}+ N_2 (\upsilon-u)~,\nn
\end{eqnarray}
where
\begin{eqnarray}
&& a_{r}=\frac{m}{-2\mathcal{E}}\Big[1\!+\!\frac{3}{2}\mathcal{E}\!+\!\frac{1}{4}\mathcal{E}^2\!+\!8\frac{m^2\mathcal{E}}{\mathcal{J}^2}\Big(1\!-\!\frac{3g^2}{8m^2}\Big)\Big]~,\nn\\
&& e_r^2=1\!+\!\frac{2\mathcal{E}\mathcal{J}^2}{m^2}\!-\!\mathcal{E}\Big(8\!+\!7\frac{\mathcal{E}\mathcal{J}^2}{m^2}\Big)
\!+\!\,\mathcal{E}^2\Big[16\frac{\mathcal{E}\mathcal{J}^2}{m^2}\!-\!20\Big(1\!-\!\frac{3g^2}{5m^2}\Big)\!-\!32\frac{m^2}{\mathcal{E}\mathcal{J}^2}
\Big(1\!-\!\frac{3g^2}{8m^2}\Big)\Big]~,\nn\\
&& e_t=e_r\Big[1\!+\!6\,\mathcal{E}\!+\!27\,\mathcal{E}^2\!-\!\frac{30m\mathcal{E}^2}{\sqrt{-2\mathcal{E}\mathcal{J}^2}}\!+\!8\frac{m^2\mathcal{E}}{\mathcal{J}^2}
\Big(1\!-\!\frac{3g^2}{8m^2}\Big)\Big]~,\nn\\
&& e_{\phi} = e_r\Big[1\!-\!2\mathcal{E}\!-\!18\frac{m^2\mathcal{E}}{\mathcal{J}^2}\Big(1\!-\!\frac{g^2}{6m^2}\Big)\Big]~,\nn\\
&& \Phi=2\pi\Big[1\!+\!3\frac{m^2}{\mathcal{J}^2}\!+\!\frac{15}{2}\frac{m^2\mathcal{E}}{\mathcal{J}^2}\!+\!\frac{105}{4}\frac{m^4}{\mathcal{J}^4}
\Big(1\!-\!\frac{6g^2}{35m^2}\Big)\Big]~,\nn\\
&&  N_0= \frac{m^2}{\mathcal{J}^2}\Big[1\!+\!2\mathcal{E}\!+\!10\frac{m^2}{\mathcal{J}^2}\Big(1\!-\!\frac{3g^2}{20m^2}\Big)\Big]
\Big(1\!+\!\frac{2\mathcal{E}\mathcal{J}^2}{m^2}\Big)^{\!\frac{1}{2}}~,\nn\\
&&  N_1= \frac{3}{8}\frac{m^4}{\mathcal{J}^4}\Big(1+\frac{2\mathcal{E}\mathcal{J}^2}{m^2}\Big)~,\nn\\
&&  N_2 =\frac{30m \,\mathcal{E}^2 }{\sqrt{-2\,\mathcal{E}\mathcal{J}^2}}~,\nn\\
&&  N_3= \frac{1}{8}\frac{m^4}{\mathcal{J}^4}\Big(1+\frac{2\mathcal{E}\mathcal{J}^2}{m^2}\Big)~,\nn\\
&& {\rm T}_u=\frac{2\pi m}{(-2\mathcal{E})^{\frac{3}{2}}}\Big(1\!-\!\frac{15}{4}\mathcal{E}\!-\!\frac{105}{32}\mathcal{E}^2
+\frac{30m\mathcal{E}^2}{\sqrt{-2\mathcal{E}\mathcal{J}^2}}\Big)~.\nn
\end{eqnarray}
In the formulations, $a_{r}$, $e_{r}$ and $u$ denote the semi-major axis, the eccentricity, the eccentric anomaly of the quasi-Keplerian motion in the PN approximation, respectively. $f$ and $\upsilon$ are two slightly different definitions of the true anomaly. ${\rm T}_u$ denotes the orbital period. 
Notice that the effects of the magnetic charge on the test particle's motion are characterized by the terms containing $g^2$ in the above formulas. 

The test particle's orbital perihelion precession can be calculated by \begin{eqnarray}
\Delta \phi \!\equiv\! \Phi-2\pi=6\pi\frac{m^2}{\mathcal{J}^2}\!+\!15\frac{\pi m^2\mathcal{E}}{\mathcal{J}^2}\!+\!\frac{105}{2}\frac{\pi m^4}{\mathcal{J}^4}
\Big(1\!-\!\frac{6g^2}{35m^2}\Big)~.\label{deltaphi}
\end{eqnarray}
When the magnetic charge of the Bardeen black hole vanishes ($g=0$), Eq.~(\ref{deltaphi}) can be simplified to the 2PN Schwarzschild periastron precession~\cite{DamourSchafer1988,MGS2004}.

Finally, it can be observed that the test particle's orbital period is independent on the the magnetic charge up to the 2PN order.

\section*{ACKNOWLEDGEMENT}
This work was supported in part by the National Natural Science Foundation of China (Grant Nos. 11973025 and 11947404).

\bibliography{Reference_20220609}

\end{document}